\documentclass{article}

\usepackage{arxiv}

\usepackage[utf8]{inputenc} % allow utf-8 input
\usepackage[T1]{fontenc}    % use 8-bit T1 fonts
\usepackage{hyperref}       % hyperlinks
\usepackage{url}            % simple URL typesetting
\usepackage{booktabs}       % professional-quality tables
\usepackage{amsfonts}       % blackboard math symbols
\usepackage{nicefrac}       % compact symbols for 1/2, etc.
\usepackage{microtype}      % microtypography
\usepackage{lipsum}		% Can be removed after putting your text content
\usepackage{graphicx}
\usepackage[sort&compress,numbers]{natbib}
\usepackage{doi}

\title{Observation of Weak Kondo Effect and Angle Dependent Magnetoresistance in Layered Antiferromagnetic V$_5$S$_8$ Single Crystal}

\author{ \href{https://orcid.org/0000-0002-8478-2948}{\includegraphics[scale=0.06]{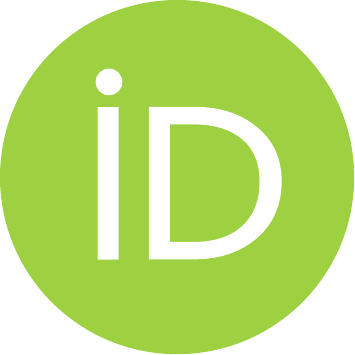}\hspace{1mm}Indrani Kar} \\
	Department of Condensed Matter and Materials Physics\\
	S. N. Bose National Centre for Basic Sciences\\
	Salt Lake, JD Block, Sector III, Bidhannagar, Kolkata-700106, India \\
	\And
	\href{https://orcid.org/0000-0001-9899-4872}{\includegraphics[scale=0.06]{orcid.pdf}\hspace{1mm}Sayan Routh} \\
    Department of Condensed Matter and Materials Physics\\
	S. N. Bose National Centre for Basic Sciences\\
	Salt Lake, JD Block, Sector III, Bidhannagar, Kolkata-700106, India \\
	\AND
	\href{https://orcid.org/0000-0001-6812-0250}{\includegraphics[scale=0.06]{orcid.pdf}\hspace{1mm}Soumya Ghorai} \\
	Department of Condensed Matter and Materials Physics\\
	S. N. Bose National Centre for Basic Sciences\\
	Salt Lake, JD Block, Sector III, Bidhannagar, Kolkata-700106, India \\
	\And
    \href{https://orcid.org/0000-0001-9861-7778}{\includegraphics[scale=0.06]{orcid.pdf}\hspace{1mm}Shubham Purwar} \\
	Department of Condensed Matter and Materials Physics\\
	S. N. Bose National Centre for Basic Sciences\\
	Salt Lake, JD Block, Sector III, Bidhannagar, Kolkata-700106, India \\
    \And
    \href{https://orcid.org/0000-0003-1258-0981}{\includegraphics[scale=0.06]{orcid.pdf}\hspace{1mm}S. Thirupathaiah*} \\
	Department of Condensed Matter and Materials Physics\\
	S. N. Bose National Centre for Basic Sciences\\
	Salt Lake, JD Block, Sector III, Bidhannagar, Kolkata-700106, India \\
	\texttt{setti@bose.res.in} \\
}

\renewcommand{\shorttitle}{Weak Kondo Effect and Angle Dependent Magnetoresistance in V$_5$S$_8$}

\hypersetup{
pdftitle={A template for the arxiv style},
pdfsubject={q-bio.NC, q-bio.QM},
pdfauthor={David S.~Hippocampus, Elias D.~Striatum},
pdfkeywords={First keyword, Second keyword, More},
}

\begin{document}
\maketitle

\begin{abstract}
The compound V$_5$S$_8$ can also be represented by V$_{1.25}$S$_2$, a transition metal dichalcogenide (TMDC) with excess V. Very few TMDCs show magnetism and/or Kondo effect. Among them, the sister compounds VSe$_2$ and VTe$_2$ are recently proved to show ferromagnetism in addition to the low-temperature resistivity upturn due to Kondo effect. In this study, we show Kondo effect in  V$_5$S$_8$ originated from the antiferromagnetic exchange interactions among the intercalated V atoms below the N$\acute{e}$el ($T_N$) temperature of 27 K. We find isotropic magnetic properties above $T_N$, while a strong magnetic anisotropy is noticed below $T_N$. In addition, below $T_N$ we find an out-of-plane ($H\parallel c$) spin-flop transition triggered at a critical field of 3.5 T  that is absent from the in-plane ($H\perp c$). Angle-dependent magnetoresistance is found to be highly anisotropic in the antiferromagnetic state.
\end{abstract}

% keywords can be removed

\section{Introduction}
The discovery of magnetism in 2-dimensional (2D) layered systems~\cite{Tokmachev2021, Dai2022, Ji2022, Li2022} has attracted a great deal of research interest for the last few years because of their versatile technological applications in spintronics~\cite{Kim2018a, Song2019, Wang2018a, Lin2020, Ye2021}. These systems possess loosely bound 2D layers connected by the van der Waals forces~\cite{Sethulakshmi2019}. By stacking~\cite{Camley1993}, twisting~\cite{Xie2022a} the monolayers, or by making the heterostructured magnetic materials~\cite{Song2018, Belim2022}, the  magnetic properties can be tuned. Though magnetism is rare in the transition metal dichalcogenides (TMDCs) in their bulk form, there exist few TMDCs showing ferromagnetic ordering mostly in low dimensions. For instance, monolayer 2$H$-NbSe$_2$~\cite{Pervin2019, Wickramaratne2020}, monolayer/bilayer VSe$_2$~\cite{Li2014a, Bonilla2018}, monolayer VTe$_2$~\cite{Fuh2016}, nano-$2H$-NbSe$_2$~\cite{Gill2017}, and atomically thin $1T$-NbS$_2$~\cite{Martino2021} show ferromagnetism. Nevertheless, VSe$_2$ is found to show ferromagnetism in its bulk phase~\cite{Li2014a}. In addition to ferromagnetism, a very few TMDCs are known to show the Kondo effect such as ZrTe$_2$~\cite{Wang2021}, VSe$_2$~\cite{Barua2017, Pandey2020}, and VTe$_2$~\cite{Ding2021, Liu2019a} due to low temperature magnetic ordering induced by the intercalated transition metal. The Kondo effect arises from the exchange interaction between localized magnetic moments of the impurities and conduction electrons in metals, leading to increase in electrical resistivity at extremely low temperatures. Thus, the doped or intercalated magnetic impurities  with unfilled $d$- or $f$-orbitals~\cite{Krellner2007, Nevidomskyy2009, Li2013, Sarkar2015} act as the scattering centres for this type of exchange interactions.

\begin{figure}[t]
  \centering
  \includegraphics[width=0.49\textwidth]{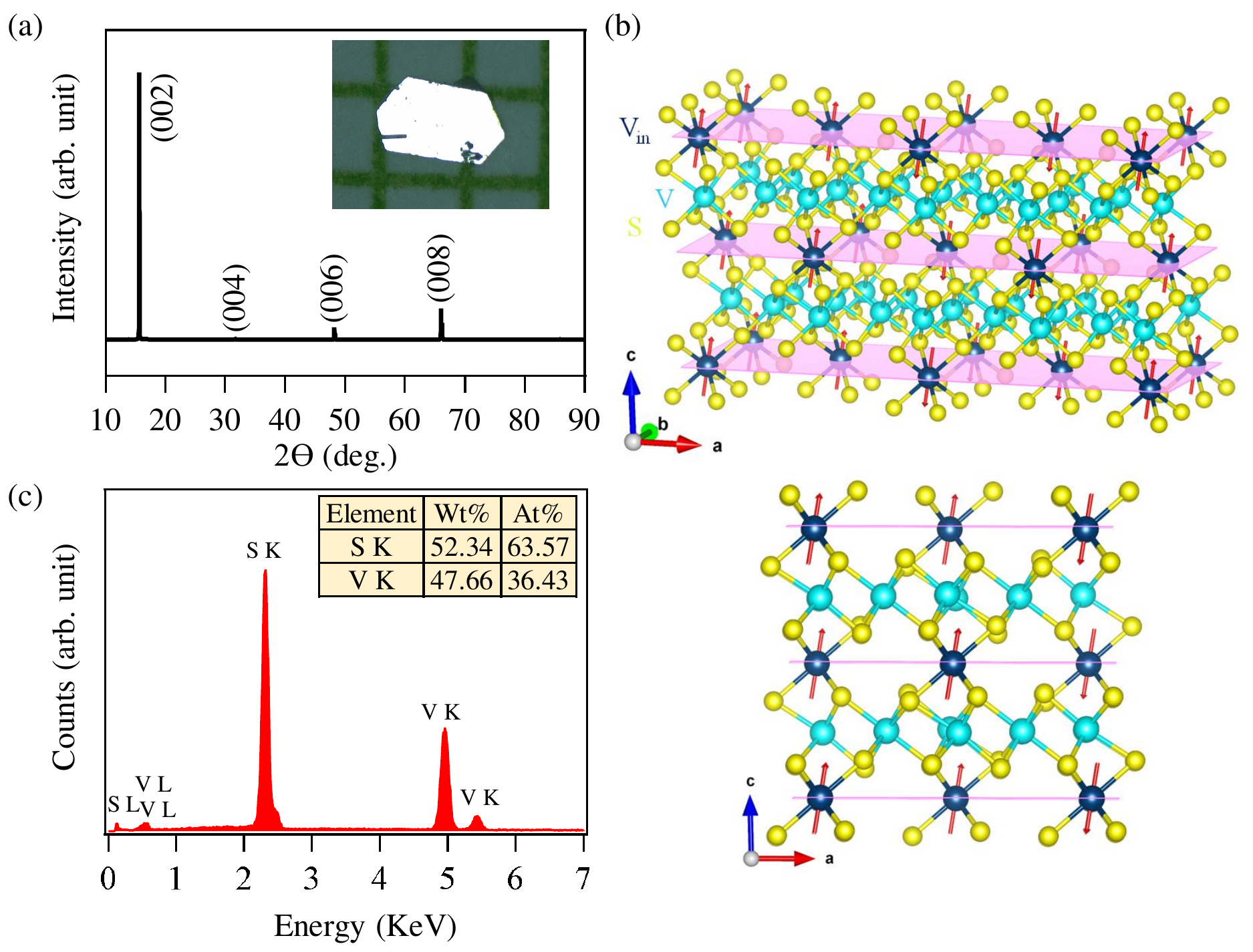}
  \caption{(a) Powder X-ray diffractogram (XRD) of  V$_5$S$_8$ single crystals. Inset is a photographic optical image of the single crystal. Top panel in (b) is the crystal structure of  V$_5$S$_8$ spanned over two unit cells. Blue color atoms are intercalated vanadium atoms in the van der Waals gap between two layers of VS$_2$. Bottom panel in (b) is the  V$_5$S$_8$ crystal structure projected onto the $ac$-plane. (c) Energy dispersive X-ray analysis (EDX) of  V$_5$S$_8$ single crystals. Inset in (c) shows atomic and weight percentages of the elements present in the compound. }
  \label{1}
\end{figure}

Recently V$_5$S$_8$, which can be expressed as V$_{1.25}$S$_2$, having 25\% excess V in the composition has been reported to possess Kondo lattice~\cite{Niu2020, Zhou2022} similar to VTe$_2$ and VSe$_2$ single crystals~\cite{Barua2017, Ding2021}. Angle dependent magnetoresistance studies and measurements of critical magnetic field at which Kondo effect vanishes have been reported for both VTe$_2$ and VSe$_2$ single crystals~\cite{Barua2017, Ding2021}, however systematic study of the Kondo effect and Magnetoresistance in the bulk V$_5$S$_8$ single crystal is still due. The ferromagnetism and Kondo effect arise in VTe$_2$ and VSe$_2$ by the intercalated V atoms~\cite{Barua2017, Pandey2020, Ding2021}. Similarly, in V$_5$S$_8$, the V atoms are intercalated between the VS$_2$ layers~\cite{Silbernagel1975}. The V intercalation in V$_5$S$_8$ seems to be completely changing the crystal structure and physical properties from the pristine VS$_2$. Precisely, the pristine VS$_2$ is a nonmagnetic metal in the $1T$ phase with a charge density wave (CDW) ordering at 305 K~\cite{Mulazzi2010, Gauzzi2014}, while V$_5$S$_8$ is an antiferromagnetic metal in the $1T^\prime$ phase without CDW ordering~\cite{DeVries1973, Silbernagel1975, Nozaki1977, Kitaoka1980, Nakanishi2000}.

\begin{figure}[ht]
  \centering
  \includegraphics[width=0.49\textwidth]{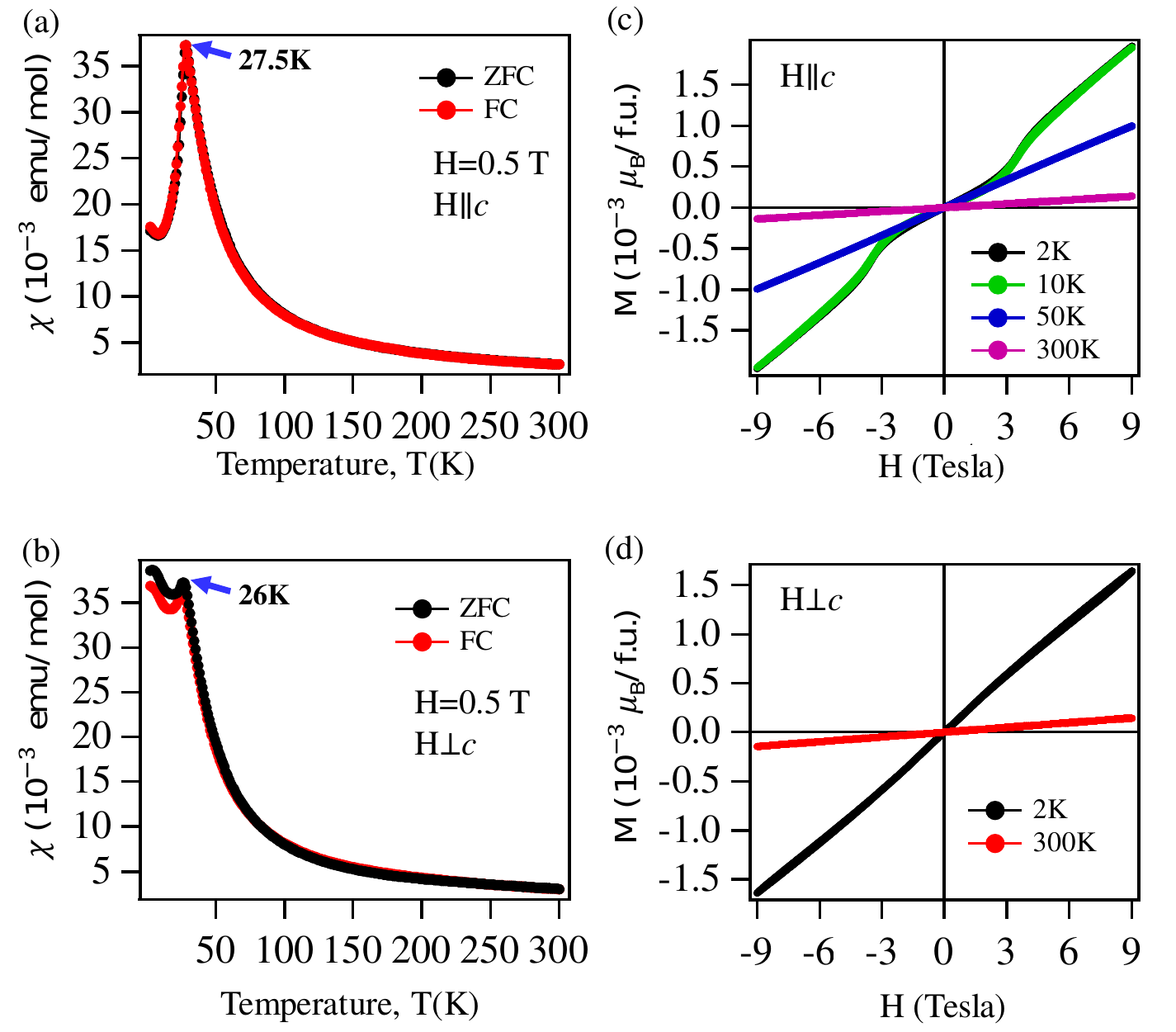}
  \caption{(a) Magnetic susceptibility plotted as a function of temperature measured with a field of 0.5 T in the FC and ZFC modes for $H\parallel c$. (b) Same as (a) but measured for $H\perp c$. (c) Magnetization isotherms $M(H)$ measured at different temperatures for $H\parallel c$. (d) Same as (c) but measured for $H\perp c$.}
  \label{2}
\end{figure}

In this work, we present electrical transport, magnetotransport, and magnetic properties studies on  V$_5$S$_8$ single crystals. Magnetization [$M(T)$] studies suggest an antiferromagnetic ordering at around 27 K, in agreement with previous reports~\cite{DeVries1973, Silbernagel1975, Nozaki1977, Kitaoka1980, Nakanishi2000}. Interestingly, down to the AFM ordering temperature ($>$27 K) the magnetic susceptibility is found to be nearly isotropic between $H\parallel c$ and $H\perp c$.  However, below the N$\acute{e}$el temperature ($<$27 K) the magnetic susceptibility instantly turns into anisotropic. Magnetization isotherms [$M(H)$] suggest a spin-flop transition at a critical field of 3.5 T for $H\parallel c$. We find an upturn in electrical resistance at low temperature with a resistance minima at 6 K,  plausibly due to the Kondo effect.  Negative magnetoresistance (MR) is noticed in the antiferromagnetic state, while it is negligible in the paramagnetic state (100 K).

\section{Experimental Details}\label{2}
Single crystals of V$_5$S$_8$ were grown by the chemical vapor transport (CVT) technique with iodine as a transport agent\cite{SCHAeFER1964}. In the first step, stoichiometric quantities of V (powder, 99.5\%, metals basis, Alfa Aesar) and S (pieces, 99.999\%, metals basis, Alfa Aesar) were mixed thoroughly and sealed in a quartz ampoule under vacuum along with pieces of crystallized iodine (crystalline, 99.99+\%, metals basis, Alfa Aesar) (2 mg/cm\textsuperscript{3}). The ampoule was kept in a three-zone tube furnace where the temperatures were set at 1000\textsuperscript{$\circ$}C for the hot-zone and 950\textsuperscript{$\circ$}C for the cold-zone.  After 5 days of reaction, we obtained shiny single crystals of V$_5$S$_8$ with a typical dimension of 2 mm$\times$1 mm at the cold-zone.

Chemical composition of the single crystals were determined by energy dispersive X-ray analysis (EDX) equipped with a scanning electron microscope (Quanta 250 FEG). Phase purity and crystal structure were confirmed by the X-ray diffraction (XRD) technique using Cu k\textsubscript{$\alpha$}-radiation (Rigaku MiniFlex II and Rigaku SmartLab 9KW). Electrical transport was done in a physical property measurement system (Quantum Design PPMS-9T) using a standard four-probe method, with the electrical current applied parallel to $ab$-plane. Magnetic field was applied at different polar angles with respect to the $c$-axis for magnetotransport measurements up to 9 T. Four copper (Cu) leads were connected to the sample by vacuum compatible silver epoxy (Epo-Tek H20E). The sample temperature was varied between 2 K and 380 K during the transport measurements. DC magnetization measurements were performed using vibrating sample magnetometer (VSM, Quantum Design PPMS-9T). Temperature dependence of the magnetization in zero-field-cooled (ZFC) and field-cooled (FC) modes were carried out under different magnetic fields within the temperature range of 2-300 K.

\section{Results and Discussions}\label{3}
\begin{figure}[htb!]
  \centering
  \includegraphics[width=0.49\textwidth]{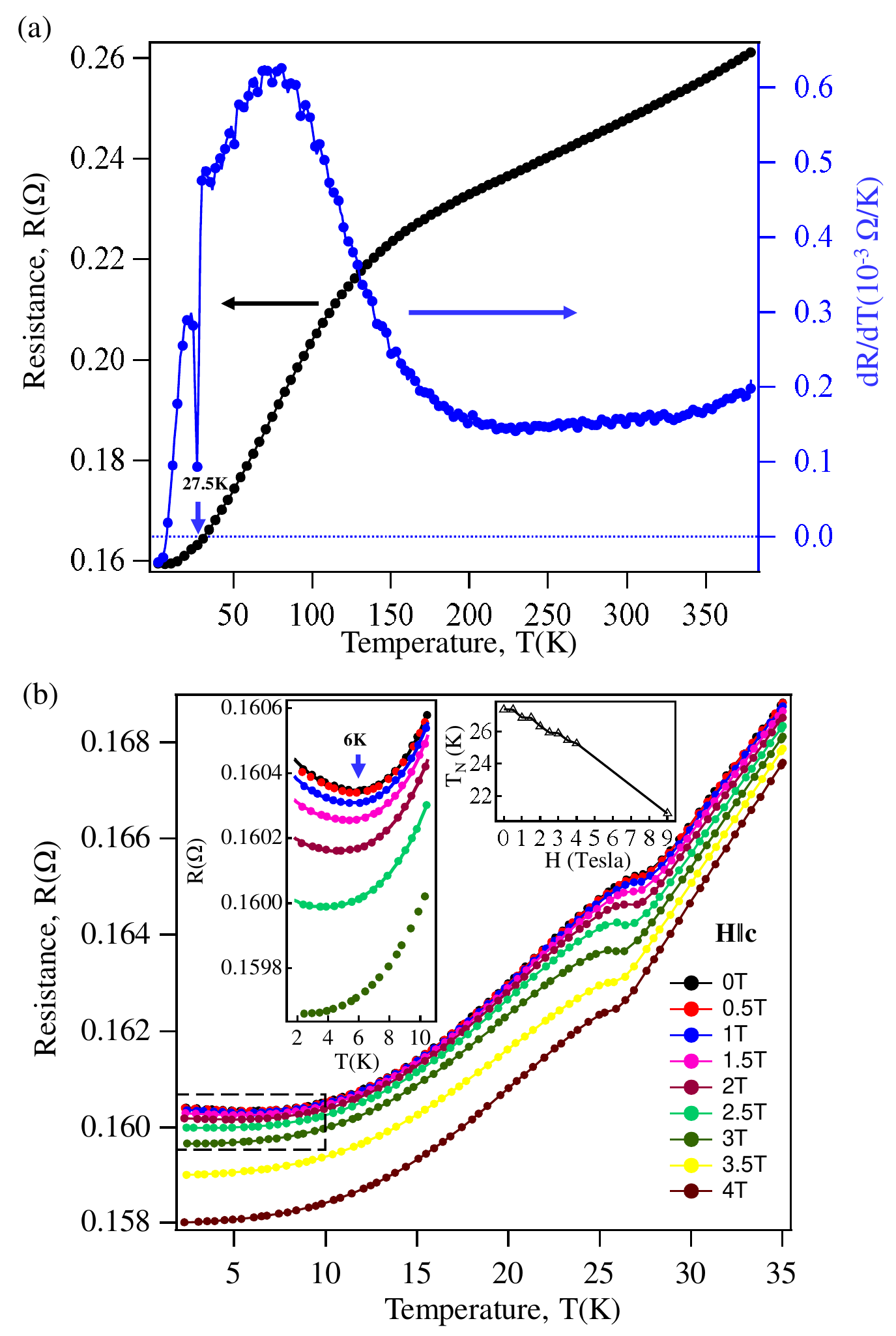}
  \caption{(a) In-plane electrical resistance and first derivative of resistance (dR/dT) plotted as a function of temperature. (b) Temperature dependent electrical resistance plotted for various magnetic fields applied parallel to the $c$-axis. Left inset in (b) is the zoomed-in image of the dash marked rectangular region. Solid lines are fitting to Eq.~\ref{eq2} at 0 T and Eq.~\ref{eq3} in presence of field. Right inset in (b) shows N$\acute{e}$el temperature $T_N$ plotted as a function of applied field.}
  \label{3}
\end{figure}

From Fig.~\ref{1}(a) sharp Bragg peaks are observed in the XRD pattern of V$_5$S$_8$ single crystal corresponding to the $(0~0~l)$ plane, suggesting that the crystal growth is along the $c$-axis.  The single crystals were in platelike shape with metallic luster as illustrated in the inset of Fig.~\ref{1}(a). Top panel in Fig.~\ref{1}(b) shows the crystal structure of V$_5$S$_8$ in which the S atoms form covalent bonding with the intercalated V atoms in the octahedral coordination in between two VS$_2$ layers. These intercalated V atoms are responsible for the distortion in the octahedral coordination of  VS$_2$, leading to antiferromagnetism with spins aligned at an angle of 10.4$^\circ$ from the $c$-axis as demonstrated in the bottom panel of Fig.~\ref{1}(b)~\cite{Oka1974, Nozaki1978, Funahashi1981}. Further, V$_5$S$_8$ crystallizes into the monoclinic structure of the space group $F12/m1$(12) with distorted $1T$ ($1T^\prime$) phase. From the EDX measurements, shown in Fig.~\ref{1}(c), we estimate actual chemical composition of the obtained single crystals to V$_{4.6}$S$_8$ which is very close to the nominal composition of V$_5$S$_8$.

Figs.~\ref{2}(a) and ~\ref{2}(b) show magnetic susceptibility ($\chi$) plotted as a function of temperature with field (H=0.5T) applied parallel ($H\parallel c$) and perpendicular ($H\perp c$) to the $c$-axis, respectively.  From Figs.~\ref{2}(a) and ~\ref{2}(b) we notice similar trends of susceptibility with respect to the temperature in going from 300 K down to 27$\pm$1 K for both $H\parallel c$ and $H\perp c$, suggesting isotropic magnetic properties between 300 and 27$\pm$1 K. But below 27$\pm$1 K, we see a sudden drop in $\chi$ for $H\parallel c$ like in an antiferromagnetic system. This observation is consistent with previous reports on these systems where an AFM order is demonstrated below $T_N$=27 K~\cite{DeVries1973, Silbernagel1975, Nozaki1977, Nozaki1978, Kitaoka1980, Funahashi1981, Nakanishi2000, Niu2017}. On the other hand, for $H\perp c$, though we find a slight drop in $\chi$  below 27$\pm$1 K, the susceptibility gets saturated upon decreasing the temperature down to 2 K. Thus, the system shows clear anisotropic magnetic properties below the N$\acute{e}$el temperature.

\begin{figure*}[htb!]
  \centering
  \includegraphics[width=0.98\textwidth]{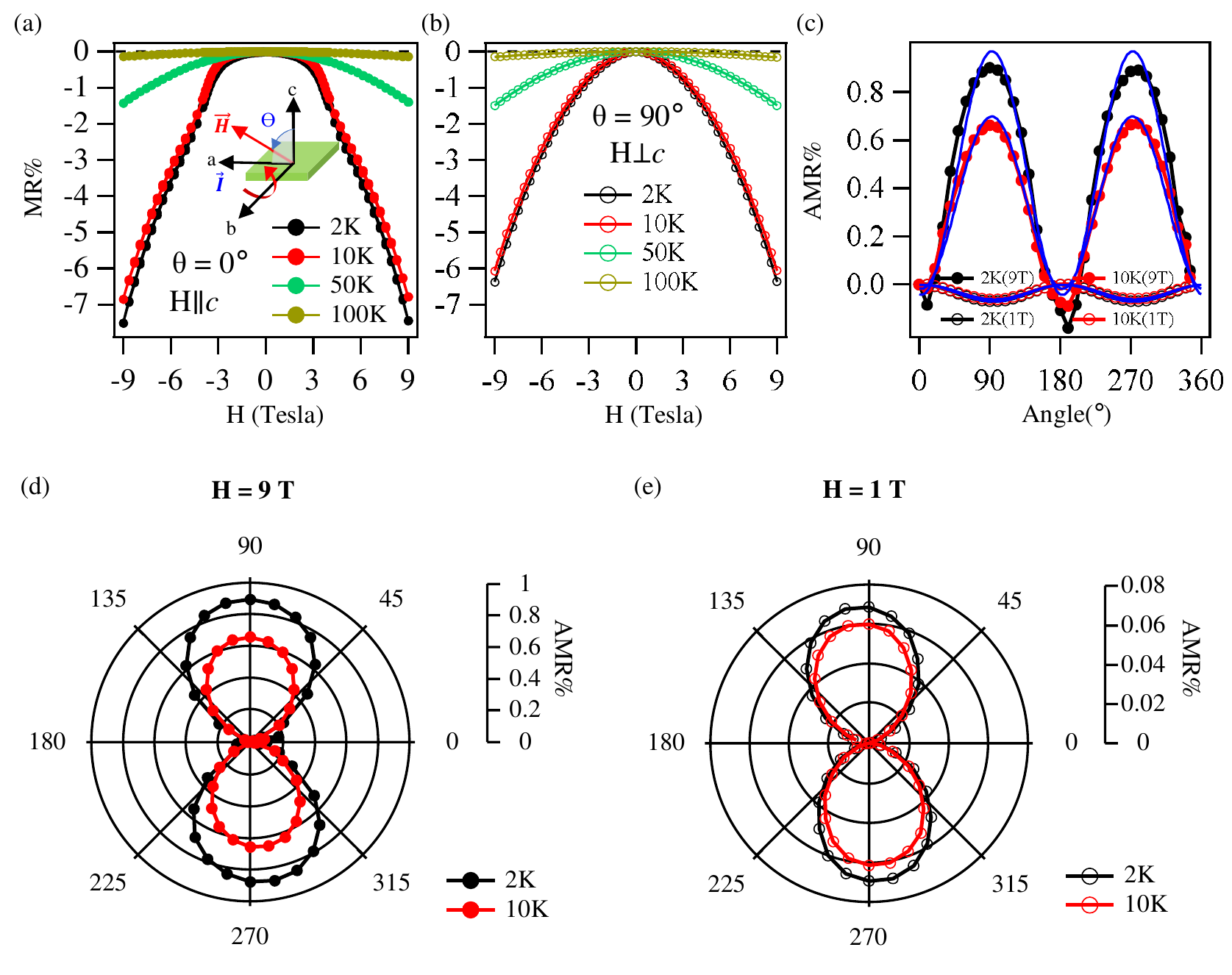}
  \caption{(a) Magnetoresistance, MR (\%),  plotted as a function of field applied parallel to the $c$-axis ($\theta=0^\circ$). (b) Same as (a) but measured for the field applied perpendicular to the $c$-axis ($\theta=90^\circ$). (c) Angle-dependent MR (\%) plotted as function of field-angle under the applied fields of 1 T and 9 T. (d) and (e) AMR (\%) of (c) plotted in the polar graph with an applied of 9T and 1T respectively.}
  \label{4}
\end{figure*}

Fig.~\ref{2}(c) depicts  magnetization isotherms [$M(H)$] measured at different temperatures with field applied parallel to the $c$-axis. While the data largely resembles the AFM type ordering at all the measured temperatures , at low temperatures (2 and 10 K) we observe spin-flop (SF) like transitions at a critical field of 3.5 T, in agreement with previous reports~\cite{Nozaki1977, Nozaki1978, Niu2017}.  Fig.~\ref{2}(d) shows $M(H)$ data measured at different temperatures with field applied perpendicular to the $c$-axis. Similar to the  $H\parallel c$ data, from $H\perp c$ also we observe AFM type ordering except that SF-transition is not observed down to the lowest possible measured temperature. This indicates that $c$-axis is the easy-axis of magnetization in V$_5$S$_8$~\cite{Oka1974, Nozaki1978, Funahashi1981}.

%similar plots as top panel with orientation of applied magnetic field as $B{\parallel}ab$. Here, in Fig.~\ref{3} (f) SF transition is destroyed from the lowest possible temperature of 2 K in similarity with the observation in MR plot in Fig.~\ref{3} (a) (right). SF transition is angle dependent and more prominent along easy axis and its neighbourhood~\cite{Xiao2022}. As $c$-axis is close to its easy axis~\cite{Oka1974, Nozaki1978, Funahashi1981} in this system, the SF transition is observed when $B \perp ab$ or $B{\parallel}c$.

\begin{table*}[htbp]
 \begin{equation}\label{eq2}
  R(T)=R_{0}+aT^2+bT^5+R_{KO}[1-\frac{\ln{(\frac{T}{T_K})}}{\sqrt{\ln^{2}{(\frac{T}{T_K})}+S(S+1)\pi^2}}]
\end{equation}
\end{table*}

\begin{table*}[htbp]
\begin{equation}\label{eq3}
R(T)=R_{0}+aT^2+bT^5+R_{KO}[1-\frac{\ln{(\frac{T}{T_K})}}{\sqrt{\ln^{2}{(\frac{T}{T_K})}+S(S+1)\pi^2}}][1-B^2(\frac{{g\mu_B}SH}{{K_B}(T+T_K)})]
\end{equation}
\end{table*}

Fig.~\ref{3}(a) depicts temperature dependent zero-field in-plane electrical resistance of V$_5$S$_8$. The resistance data suggest a metallic behaviour at low temperature, consistent with previous studies~\cite{Niu2017}. We further observe a hump in the resistance at around 27 K due to antiferromagnetic ordering~\cite{Niu2017}.  To better understand the electrical resistance,  we performed first derivative of the resistance with respect to temperature ($dR/dT$) as shown in Fig.~\ref{3}(a).  Again, we clearly see a drastic change in the slope ($dR/dT$) at around 27 K corresponding to the resistance hump.  However from a careful observation of $dR/dT$, we find that below 6 K it becomes negative as the resistance increases with decreasing temperature (resistance upturn). To understand the mechanism of resistance upturn, we measured it at various applied magnetic fields within the temperature range of 2-35 K as shown in Fig.~\ref{3}(b). Foremost, we observe from Fig.~\ref{3}(b) that the peak temperature of hump-like structure decreases with increasing applied field from 27 K at 0 T to 21 K at 9 T ($R(T)$ at 9 T is not shown) as shown in the right inset of Fig.~\ref{3}(b). This observation reaffirms the AFM ordering in this system. Secondly, as shown in the left inset of Fig.~\ref{3}(b), the resistance upturn with a minima ($T_m$) at 6 K disappears above 2.5 T.

In principle, there are mainly three reasons behind the resistance upturn. They are i) electron-electron interaction (EEI)~\cite{Lee1985}, ii) weak localization (WL)~\cite{Altshuler1980}, and iii) Kondo effect~\cite{Kondo1964}. Among these three, EEI shows positive MR~\cite{Ghosh2017} whereas negative MR is observed in Kondo effect~\cite{Wang2021, Barua2017, Ding2021} and WL~\cite{Lu2015, Wang2021}.  In addition, by increasing applied magnetic field,  the resistance minima can be suppressed in Kondo effect and WL whereas applied field strength has no effect in the case of EEI. Thus, as can be seen from inset of Fig~\ref{3}(b), the resistance minima is suppressed at 2.5 T.  Further, we demonstrate below a negative magnetoresistance in V$_5$S$_8$. Thus, we can rule out EEI as the origin of resistance upturn. Traditionally,  WL and weak antilocalization (WAL) are present in low dimensional systems~\cite{Lu2014} such as thin films~\cite{Niu2016}, nanowires~\cite{Sett2017} due to the higher probability of scattering rates resulting into quantum interference~\cite{Sasmal2020}.  Moreover, it is well known phenomenon that a small amount of magnetic field can destroy the quantum interference and lead to a cusp-like positive or negative magnetoresistance around zero-field as a result of WL or WAL~\cite{Lu2015, Niu2016, Sasmal2020, Laha2021, Wang2021}. On the other hand, the Kondo effect exhibits negative MR having quadratic dependence on B in low field region~\cite{Katayama1967, Barua2017, Liu2019a, Wang2021, Ding2021}. As we show below,  MR  of  V$_5$S$_8$ depends quadratically on the field. Importantly, the cusp-like feature is not observed from our MR measurements of  V$_5$S$_8$. Also, the conductivity ($\sigma(T)$) is not fitted well (not shown) with the WL equation $\sigma= \sigma_0+kT^{p/2}$ where $k=\frac{2e^2}{ah\pi^2}$ and the factor $a$ is defined in terms of Thoules inelastic collision length ($L_{Th}$) as $L_{Th} =aT^{-p/2}$ for p=2, 3, and 3/2~\cite{Lee1985, Sett2017}.  Thus, the WL effect also can be excluded from the discussion. Finally, we conclude that the Kondo effect is mainly causing the low temperature resistance upturn in V$_5$S$_8$. This observation is consistent with previous reports on V$_5$S$_8$ single crystal~\cite{Niu2020} and nanoflake~\cite{Zhou2022}. In case of V$_5$S$_8$ nanoflake~\cite{Zhou2022}, the upturn is clearly observed up to magnetic field of 12 T whereas in our sample the upturn is suppressed at 2.5 T. This indicates that the Kondo effect in V$_5$S$_8$ is much robust to magnetic field in lower dimension than in the bulk. Note that the magnetic field dependence of upturn is not clearly discussed in previous study on bulk V$_5$S$_8$ single crystal~\cite{Niu2020}.

Having confirmed the Kondo effect in V$_5$S$_8$, the resistance data measured without and with magnetic fields (up to $H$=2.5 T) are fitted using the Eqs.~\ref{eq2} and ~\ref{eq3}, respectively, as shown in the inset of Fig.~\ref{3}(b)~\cite{Liu2019a}. In Eq.~\ref{eq2}, first term ($R_0$) is the residual resistance, second term ($aT^2$) represents the Fermi-liquid contribution, third term ($bT^5$) represents the electron-phonon contribution, and the fourth term is the Kondo resistance described by the Hamann expression~\cite{Hamann1967, Liu2019a}. In the fourth term, $R_{KO}$ is the temperature-independent Kondo resistance, $T_K$ is the Kondo temperature, and $S$ is the total spin of the magnetic impurity. The resistance curves are best fitted with $T_K$ = 6 K and S=1 since V\textsuperscript{3+} ($3d^2$) ions carry the localized magnetic moment in this system as reported earlier~\cite{DeVries1973, Silbernagel1975, Kitaoka1980, Funahashi1981}. In Eq.~\ref{eq3},  the Hamann term is modified using quantum Brillouin function, $B(x)=({\frac{2S+1}{2S}}){\coth{{\frac{2S+1}{2S}}}x}-{\frac{1}{2S}}{\coth{{\frac{1}{2S}}}x}$ . Here, $g$ is the Land$\grave{e}$ $g$-factor, $\mu_B$ is the Bohr magneton, and $k_B$ is the Boltzmann constant~\cite{Liu2019a, Barua2017, Kondo1999, Cao2017}.

\begin{table}
\footnotesize
  \centering
    \begin{tabular}{|p{.04\textwidth}| p{.06\textwidth}| p{.08\textwidth}| p{.08\textwidth}| p{.09\textwidth}|}
    \hline
     \hfil H (T) & \hfil $R_0$ ($\Omega$) & \hfil $a$ (${\mu}{\Omega}-K^{-2}$) & \hfil $b$ (${n}{\Omega}-K^{-5}$) & \hfil $R_{KO}$ ($\Omega$) \\
  \hline

  \hfil 0 & \hfil 0.1598 & \hfil 0.926 & \hfil 1.996 & \hfil 0.000536 \\
  \hline

  \hfil 1 & \hfil 0.1596 & \hfil 1.248 & \hfil 1.863 & \hfil 0.000616 \\
  \hline

  \hfil 1.5 & \hfil 0.1597 & \hfil 1.209 & \hfil 1.890 & \hfil 0.000510 \\
  \hline

  \hfil 2 & \hfil 0.1597 & \hfil 1.621 & \hfil 1.679 & \hfil 0.000438 \\
  \hline

  \hfil 2.5 & \hfil 0.1596 & \hfil 2.223 & \hfil 1.453 & \hfil 0.000359 \\
  \hline
  \end{tabular}
  \caption{Kondo fitting parameters}\label{T1}
\end{table}

%Table~\ref{T1} represents the obtained values of parameters. The obtained value of $T_w$ is comparable with its isostructural \ch{VTe2}~\cite{Ding2021} but higher than the obtained value for isovalent \ch{VSe2} with $T_w \approx 1.2$~\cite{Barua2017} and \ch{ZrTe2} with $T_w \approx 2.2$~\cite{Wang2021}. The higher value of $T_w$ indicates stronger RKKY interaction induced by indirect exchange coupling of nuclear magnetic moments by conduction electrons~\cite{Ruderman1954}.

Figs.~\ref{4}(a) and ~\ref{4}(b)  depict magnetoresistance, $MR(\%)=\frac{R(H)-R(0)}{R(0)} \times 100\%$,  plotted as a function of field applied parallel ($\theta=0^\circ$) and perpendicular ($\theta=90^\circ$) to the $c$-axis, respectively. We notice negative MR at low temperatures (2, 10, and 50 K), but it is negligible at 100 K. Mostly, the negative MR is observed in the Kondo state of the Kondo systems such as in La$_{1-x}$Pr$_{x}$NiO$_{3-\delta}$~\cite{Hien-Hoang2021} and VTe$_2$~\cite{Ding2021}, and in the AFM state of the AFM systems such as in EuTe$_2$~\cite{Yin2020} and FeNbTe$_2$~\cite{Qi2022, Bai2019}. However in the studied system of  V$_5$S$_8$, negative MR is noticed above and below the  Kondo temperature (6 K) and  above and below the N$\acute{e}$el temperature (27$\pm$1 K).  Also, in agreement with a previous study on this system~\cite{Nozaki1977, Nozaki1978, Niu2017}, we observe a change in MR(\%) below the N$\acute{e}$el temperature when measured with the field parallel to the $c$-axis($\theta=0^\circ$) due to  spin-flop (SF) transition at a critical field of 3.5 T. However, the same is not observed in MR(\%) when measured with field perpendicular to the $c$-axis($\theta=90^\circ$). Overall, the field dependence of MR is found to be quadratic.

%La$_{1-x}$Pr$_{x}$NiO$_{3-\delta}$~\cite{Hien-Hoang2021} and \ch{VTe2}~\cite{Ding2021}. Negative MR in AFM state is also observed in other antiferromagnetic compounds like EuTe$_2$~\cite{Yin2020}, FeNbTe$_2$~\cite{Qi2022, Bai2019}. The field dependence of MR is quadratic as explained in Boltzmann’s theory for standard metals~\cite{Mitra2019} and diminishes at higher temperatures associated with thermal scattering. A SF transition observed at 3.5 T when field applied at polar angles of 0$^\circ$, parallel to $c$-axis or perpendicular to $ab$-plane at 2 K and 10 K similar to previous reports~\cite{Nozaki1977, Nozaki1978, Niu2017}. This transition is absent when field applied at polar angles of 90$^\circ$, perpendicular to $c$-axis or parallel to $ab$-plane which again confirms that it is a SF transition~\cite{Xiao2022}. However, in contrast to the SF transitions in many other materials~\cite{Kimura2003, Chaudhury2009} any appreciable hysteresis is not observed from our data as mentioned earlier in this system~\cite{Niu2017} and in other systems~\cite{Oh2014}.

To explore further on the MR anisotropy, we performed angle dependent ($\theta$) MR measurements (AMR), $AMR(\%)= \frac{R(\theta^\circ)-R(0^\circ)}{R(0^\circ)} \times 100\%$,  at various temperatures under 1T and 9T applied fields as shown in Fig.~\ref{4}(c) by varying $\theta$ between $0^\circ$ and $360^\circ$.   Solid lines in  Fig.~\ref{4}(c) are the fits with equation $AMR(\theta)=C+{\alpha}Cos {2(\theta+\phi)}$, where C, $\alpha$ are constants and $\phi$ is the phase~\cite{Jovanovifmmodecuteclseci2010, Singha2018, Wang2021a}. From the AMR data, we observe that the oscillations exhibit large amplitudes under the magnetic field of 9 T measured at 2 K and 10 K with a twofold symmetry having maximal MR values for $H\perp c$ and minimal MR values for $H\parallel c$~\cite{McGuire1975, Qi2022}. On the other hand,  under the magnetic field of 1 T,  measured at 2 K and 10 K,  the twofold symmetry still survives, but the oscillation magnitude gets significantly reduced. For a better representation, the AMR data of Fig.~\ref{4}(c) are plotted into the polar graphs as shown in Figs.~\ref{4}(d) and ~\ref{4}(e) for the fields of 9T and 1T,  respectively. Thus, the two-fold asymmetry of MR measured at both 1 T and 9 T is clearly visible from Fig.~\ref{4}(c).

\section{Conclusions}\label{4}
In conclusion, we have systematically studied the electrical transport, magnetotransport, and magnetic properties of the V intercalated transition metal dichalcogenide V$_5$S$_8$. In this study, we show Kondo effect in V$_5$S$_8$ originated from the antiferromagnetic exchange interactions among the intercalated V atoms below the N$\acute{e}$el temperature. We find isotropic magnetic properties above $T_N$, while a strong magnetic anisotropy is noticed below $T_N$. In addition, below $T_N$ we find an out-of-plane ($H\parallel c$) spin-flop transition triggered at a critical field of 3.5 T  that is absent for $H\perp c$. Negative magnetoresistance (MR) is noticed in the antiferromagnetic state, while it is negligible in the paramagnetic state. Angle-dependent magnetoresistance is found to be highly anisotropic in the antiferromagnetic state ($<$27 K).

%When the system enters AFM state the resistance increases slightly could be due to small band gap opening in the Fermi surface. As field increases the Fermi surface modified and decrease the gap. Hence, negative MR observed in this antiferromagnetic sample could be attributed to the change of the Fermi surface as discussed in \ch{EuTe2}~\cite{Yin2020}, EuSn$_2$S$_2$~\cite{Chen2020a}.

\section{Acknowledgements}\label{5}
S.T. acknowledges financial support by the Department of Science and Technology (DST) through the grant no. SRG/2020/000393. This work is partly supported by the TRC project at the SNBNCBS.

\bibliographystyle{unsrt}
\bibliography{V5S8}  %%% Uncomment this line and comment out the ``thebibliography'' section below to use the external .bib file (using bibtex) .

\begin{thebibliography}{10}

\bibitem{Tokmachev2021}
Andrey~M. Tokmachev, Dmitry~V. Averyanov, Alexander~N. Taldenkov, Ivan~S.
  Sokolov, Igor~A. Karateev, Oleg~E. Parfenov, and Vyacheslav~G. Storchak.
\newblock {Two-Dimensional Magnets beyond the Monolayer Limit}.
\newblock {\em {ACS nano}}, 15(7):12034--12041, 2021.

\bibitem{Dai2022}
Hongwei Dai, Menghao Cai, Qinghua Hao, Qingbo Liu, Yuntong Xing, Hongjing Chen,
  Xiaodie Chen, Xia Wang, Hua-Hua Fu, and Junbo Han.
\newblock {Nonlocal Manipulation of Magnetism in an Itinerant Two-Dimensional
  Ferromagnet}.
\newblock {\em {ACS} Nano}, 16(8):12437--12444, jul 2022.

\bibitem{Ji2022}
Wei Ji.
\newblock {2D Magnetism - Materials, Devices, and Applications Forum in ACS
  Applied Electronic Materials}.
\newblock {\em ACS Applied Electronic Materials}, 4(7):3166--3167, 2022.

\bibitem{Li2022}
Yang Li, Baishun Yang, Shengnan Xu, Bing Huang, and Wenhui Duan.
\newblock {Emergent Phenomena in Magnetic Two-Dimensional Materials and van der
  Waals Heterostructures}.
\newblock {\em ACS Applied Electronic Materials}, 4(7):3278--3302, 2022.

\bibitem{Kim2018a}
Hyun~Ho Kim, Bowen Yang, Tarun Patel, Francois Sfigakis, Chenghe Li, Shangjie
  Tian, Hechang Lei, and Adam~W. Tsen.
\newblock {One million percent tunnel magnetoresistance in a magnetic van der
  Waals heterostructure}.
\newblock {\em Nano Letters}, 18(8):4885--4890, 2018.

\bibitem{Song2019}
Tiancheng Song, Matisse Wei-Yuan Tu, Caitlin Carnahan, Xinghan Cai, Takashi
  Taniguchi, Kenji Watanabe, Michael~A. McGuire, David~H. Cobden, Di~Xiao, Wang
  Yao, et~al.
\newblock {Voltage control of a van der Waals spin-filter magnetic tunnel
  junction}.
\newblock {\em Nano Letters}, 19(2):915--920, 2019.

\bibitem{Wang2018a}
Zhe Wang, Deepak Sapkota, Takashi Taniguchi, Kenji Watanabe, David Mandrus, and
  Alberto~F. Morpurgo.
\newblock {Tunneling spin valves based on Fe$_3$GeTe$_2$/hBN/Fe$_3$GeTe$_2$ van
  der Waals heterostructures}.
\newblock {\em Nano letters}, 18(7):4303--4308, 2018.

\bibitem{Lin2020}
Hailong Lin, Faguang Yan, Ce~Hu, Quanshan Lv, Wenkai Zhu, Ziao Wang, Zhongming
  Wei, Kai Chang, and Kaiyou Wang.
\newblock {Spin-valve effect in Fe$_3$GeTe$_2$/MoS$_2$/Fe$_3$GeTe$_2$ van der
  Waals heterostructures}.
\newblock {\em ACS Applied Materials \& Interfaces}, 12(39):43921--43926, 2020.

\bibitem{Ye2021}
Haoshen Ye, Yijie Zhu, Dongmei Bai, Junting Zhang, Xiaoshan Wu, and Jianli
  Wang.
\newblock {Spin valve effect in VN/GaN/VN van der Waals heterostructures}.
\newblock {\em {Phys. Rev. B}}, 103:035423, Jan 2021.

\bibitem{Sethulakshmi2019}
N.~Sethulakshmi, Avanish Mishra, P.~M. Ajayan, Yoshiyuki Kawazoe, Ajit~K. Roy,
  Abhishek~K. Singh, and Chandra~Sekhar Tiwary.
\newblock Magnetism in two-dimensional materials beyond graphene.
\newblock {\em Materials Today}, 27:107--122, jul 2019.

\bibitem{Camley1993}
R.~E. Camley and R.~L. Stamps.
\newblock Magnetic multilayers: spin configurations, excitations and giant
  magnetoresistance.
\newblock {\em Journal of Physics: Condensed Matter}, 5(23):3727, 1993.

\bibitem{Xie2022a}
Hongchao Xie, Xiangpeng Luo, Gaihua Ye, Zhipeng Ye, Haiwen Ge, Suk~Hyun Sung,
  Emily Rennich, Shaohua Yan, Yang Fu, Shangjie Tian, et~al.
\newblock Twist engineering of the two-dimensional magnetism in double bilayer
  chromium triiodide homostructures.
\newblock {\em Nature Physics}, 18(1):30--36, 2022.

\bibitem{Song2018}
Tiancheng Song, Xinghan Cai, Matisse Wei-Yuan Tu, Xiaoou Zhang, Bevin Huang,
  Nathan~P. Wilson, Kyle~L. Seyler, Lin Zhu, Takashi Taniguchi, Kenji Watanabe,
  et~al.
\newblock {Giant tunneling magnetoresistance in spin-filter van der Waals
  heterostructures}.
\newblock {\em {Science}}, 360(6394):1214--1218, 2018.

\bibitem{Belim2022}
Sergey~V. Belim, Igor~V. Bychkov, Ivan Maltsev, Dmitry~A. Kuzmin, and
  Vladimir~G. Shavrov.
\newblock {Tuning of 2D magnets Curie temperature via substrate}.
\newblock {\em Journal of Magnetism and Magnetic Materials}, 541:168553, 2022.

\bibitem{Pervin2019}
Rukshana Pervin, Manikandan Krishnan, Arumugam Sonachalam, and Parasharam~M.
  Shirage.
\newblock Coexistence of superconductivity and ferromagnetism in defect-induced
  {NbSe$_2$} single crystals.
\newblock {\em Journal of Materials Science}, 54(18):11903--11912, jun 2019.

\bibitem{Wickramaratne2020}
Darshana Wickramaratne, Sergii Khmelevskyi, Daniel~F. Agterberg, and I.~I.
  Mazin.
\newblock {Ising Superconductivity and Magnetism in {NbSe$_{2}$}}.
\newblock {\em {Phys. Rev. X}}, 10:041003, Oct 2020.

\bibitem{Li2014a}
Fengyu Li, Kaixiong Tu, and Zhongfang Chen.
\newblock {Versatile Electronic Properties of {VSe$_2$} Bulk, Few-Layers,
  Monolayer, Nanoribbons, and Nanotubes: A Computational Exploration}.
\newblock {\em The Journal of Physical Chemistry C}, 118(36):21264--21274, sep
  2014.

\bibitem{Bonilla2018}
Manuel Bonilla, Sadhu Kolekar, Yujing Ma, Horacio~Coy Diaz, Vijaysankar
  Kalappattil, Raja Das, Tatiana Eggers, Humberto~R. Gutierrez, Manh-Huong
  Phan, and Matthias Batzill.
\newblock {Strong room-temperature ferromagnetism in {VSe$_2$} monolayers on
  van der Waals substrates}.
\newblock {\em Nature Nanotechnology}, 13(4):289--293, feb 2018.

\bibitem{Fuh2016}
Huei-Ru Fuh, Ching-Ray Chang, Yin-Kuo Wang, Richard F.~L. Evans, Roy~W.
  Chantrell, and Horng-Tay Jeng.
\newblock {Newtype single-layer magnetic semiconductor in transition-metal
  dichalcogenides {VX$_2$} (X{\hspace{0.167em}}={\hspace{0.167em}}S, Se and
  Te)}.
\newblock {\em Scientific Reports}, 6(1):32625, sep 2016.

\bibitem{Gill2017}
Raminder Gill.
\newblock Superconductivity and ferromagnetism in nanomaterial {NbSe}2.
\newblock In {\em {AIP} Conference Proceedings}. Author(s), 2017.

\bibitem{Martino2021}
Edoardo Martino, Carsten Putzke, Markus König, Philip J.~W. Moll, Helmuth
  Berger, David LeBoeuf, Maxime Leroux, Cyril Proust, Ana Akrap, Holm Kirmse,
  Christoph Koch, ShengNan Zhang, QuanSheng Wu, Oleg~V. Yazyev,
  L{\'{a}}szl{\'{o}} Forr{\'{o}}, and Konstantin Semeniuk.
\newblock {Unidirectional Kondo scattering in layered NbS$_2$}.
\newblock {\em npj 2D Materials and Applications}, 5(1):86, nov 2021.

\bibitem{Wang2021}
Yihao Wang, Changzheng Xie, Junbo Li, Zan Du, Liang Cao, Yuyan Han, Lin Zu,
  Hongchao Zhang, Huamin Zhu, Xueying Zhang, Yimin Xiong, and Weisheng Zhao.
\newblock {Weak Kondo effect in the monocrystalline transition metal
  dichalcogenide ZrTe$_2$}.
\newblock {\em {Phys. Rev. B}}, 103:174418, May 2021.

\bibitem{Barua2017}
Sourabh Barua, M.~Ciomaga Hatnean, M.~R. Lees, and G.~Balakrishnan.
\newblock {Signatures of the Kondo effect in {VSe$_2$}}.
\newblock {\em Scientific Reports}, 7(1):10964, sep 2017.

\bibitem{Pandey2020}
Juhi Pandey and Ajay Soni.
\newblock {Electron-phonon interactions and two-phonon modes associated with
  charge density wave in single crystalline 1T-{VSe$_2$}}.
\newblock {\em Physical Review Research}, 2(3):033118, jul 2020.

\bibitem{Ding2021}
Xiaxin Ding, Jie Xing, Gang Li, Luis Balicas, Krzysztof Gofryk, and Hai-Hu Wen.
\newblock {Crossover from Kondo to Fermi-liquid behavior induced by high
  magnetic field in 1T-{VTe$_2$} single crystals}.
\newblock {\em Physical Review B}, 103(12):125115, mar 2021.

\bibitem{Liu2019a}
Hongtao Liu, Yunzhou Xue, Jin-An Shi, Roger~A. Guzman, Panpan Zhang, Zhang
  Zhou, Yangu He, Ce~Bian, Liangmei Wu, Ruisong Ma, Jiancui Chen, Jiahao Yan,
  Haitao Yang, Cheng-Min Shen, Wu~Zhou, Lihong Bao, and Hong-Jun Gao.
\newblock {Observation of the Kondo Effect in Multilayer Single-Crystalline
  {VTe$_2$} Nanoplates}.
\newblock {\em Nano Letters}, 19(12):8572--8580, nov 2019.

\bibitem{Krellner2007}
C.~Krellner, N.~S. Kini, E.~M. Br\"uning, K.~Koch, H.~Rosner, M.~Nicklas,
  M.~Baenitz, and C.~Geibel.
\newblock {CeRuPO: A rare example of a ferromagnetic Kondo lattice}.
\newblock {\em {Phys. Rev. B}}, 76:104418, Sep 2007.

\bibitem{Nevidomskyy2009}
Andriy~H. Nevidomskyy and P.~Coleman.
\newblock {Kondo Resonance Narrowing in $d$- and $f$-Electron Systems}.
\newblock {\em {Phys. Rev. Lett.}}, 103:147205, Oct 2009.

\bibitem{Li2013}
Yongfeng Li, Rui Deng, Weinan Lin, Yufeng Tian, Haiyang Peng, Jiabao Yi, Bin
  Yao, and Tom Wu.
\newblock {Electrostatic tuning of Kondo effect in a rare-earth-doped
  wide-band-gap oxide}.
\newblock {\em {Phys. Rev. B}}, 87:155151, Apr 2013.

\bibitem{Sarkar2015}
T.~P. Sarkar, K.~Gopinadhan, M.~Motapothula, S.~Saha, Z.~Huang, S.~Dhar,
  A.~Patra, W.~M. Lu, F.~Telesio, I.~Pallecchi, Ariando, D.~Marr{\'{e}}, and
  T.~Venkatesan.
\newblock {Unexpected observation of spatially separated Kondo scattering and
  ferromagnetism in Ta alloyed anatase {TiO}2 thin films}.
\newblock {\em Scientific Reports}, 5(1):13011, aug 2015.

\bibitem{Niu2020}
Jingjing Niu, Wenjie Zhang, Zhilin Li, Sixian Yang, Dayu Yan, Shulin Chen,
  Zhepeng Zhang, Yanfeng Zhang, Xinguo Ren, Peng Gao, Youguo Shi, Dapeng Yu,
  and Xiaosong Wu.
\newblock {Intercalation of van der Waals layered materials: A route towards
  engineering of electron correlation}.
\newblock {\em Chinese Physics B}, 29(9):097104, sep 2020.

\bibitem{Zhou2022}
Zhang Zhou, Xiaoxu Zhao, Liangmei Wu, Hongtao Liu, Jiancui Chen, Chuanyin Xi,
  Zhaosheng Wang, Enke Liu, Wu~Zhou, Stephen~J. Pennycook, Sokrates~T.
  Pantelides, Xiao-Guang Zhang, Lihong Bao, and Hong-Jun Gao.
\newblock {Dimensional crossover in self-intercalated antiferromagnetic
  V$_5$S$_8$ nanoflakes}.
\newblock {\em Physical Review B}, 105(23):235433, jun 2022.

\bibitem{Silbernagel1975}
B.~G. Silbernagel, R.~B. Levy, and F.~R. Gamble.
\newblock {Magnetic properties of V$_5$S$_8$: An NMR study}.
\newblock {\em {Phys. Rev. B}}, 11:4563--4570, Jun 1975.

\bibitem{Mulazzi2010}
M.~Mulazzi, A.~Chainani, N.~Katayama, R.~Eguchi, M.~Matsunami, H.~Ohashi,
  Y.~Senba, M.~Nohara, M.~Uchida, H.~Takagi, and S.~Shin.
\newblock {Absence of nesting in the charge-density-wave system 1T-{VS$_2$} as
  seen by photoelectron spectroscopy}.
\newblock {\em Physical Review B}, 82(7), aug 2010.

\bibitem{Gauzzi2014}
A.~Gauzzi, A.~Sellam, G.~Rousse, Y.~Klein, D.~Taverna, P.~Giura, M.~Calandra,
  G.~Loupias, F.~Gozzo, E.~Gilioli, F.~Bolzoni, G.~Allodi, R.~De Renzi, G.~L.
  Calestani, and P.~Roy.
\newblock {Possible phase separation and weak localization in the absence of a
  charge-density wave in single-phase1T-{VS}2}.
\newblock {\em Physical Review B}, 89(23), jun 2014.

\bibitem{DeVries1973}
A.~B. {De Vries} and C.~Haas.
\newblock Magnetic susceptibility and nuclear magnetic resonance of vanadium
  sulfides.
\newblock {\em Journal of Physics and Chemistry of Solids}, 34(4):651--659,
  1973.

\bibitem{Nozaki1977}
H.~Nozaki and Y.~Ishizawa.
\newblock {An evidence of spin flopping in V$_5$S$_8$ by magnetoresistance
  experiments}.
\newblock {\em Physics Letters A}, 63(2):131--132, 1977.

\bibitem{Kitaoka1980}
Yoshio Kitaoka and Hiroshi Yasuoka.
\newblock {NMR investigations on the spin fluctuations in itinerant
  antiferromagnets. II. V$_3$S$_4$ and V$_5$S$_8$}.
\newblock {\em Journal of the Physical Society of Japan}, 48(6):1949--1956,
  1980.

\bibitem{Nakanishi2000}
M.~Nakanishi, K.~Yoshimura, K.~Kosuge, T.~Goto, T.~Fujii, and J.~Takada.
\newblock {Anomalous field-induced magnetic transitions in V$_5$X$_8$
  (X=S,Se)}.
\newblock {\em Journal of Magnetism and Magnetic Materials}, 221(3):301--306,
  2000.

\bibitem{SCHAeFER1964}
H.~SCHÄFER.
\newblock {Chemical Transport Processes as an Aid in Preparative Chemistry.
  Combination of Transport Reactions with Other Processes}.
\newblock In {\em Chemical Transport Reactions}, pages 115--131. Elsevier,
  1964.

\bibitem{Oka1974}
Y.~Oka, K.~Kosuge, and S.~Kachi.
\newblock {Magnetic properties of V$_5$S$_8$}.
\newblock {\em Physics Letters A}, 50(4):311--312, 1974.

\bibitem{Nozaki1978}
H.~Nozaki, M.~Umehara, Y.~Ishizawa, M.~Saeki, T.~Mizoguchi, and M.~Nakahira.
\newblock {Magnetic properties of V$_5$S$_8$ single crystals}.
\newblock {\em Journal of Physics and Chemistry of Solids}, 39(8):851--858,
  1978.

\bibitem{Funahashi1981}
Satoru Funahashi, Hiroshi Nozaki, and Isao Kawada.
\newblock {Magnetic structure of V$_5$S$_8$}.
\newblock {\em Journal of Physics and Chemistry of Solids}, 42(11):1009--1013,
  1981.

\bibitem{Niu2017}
Jingjing Niu, Baoming Yan, Qingqing Ji, Zhongfan Liu, Mingqiang Li, Peng Gao,
  Yanfeng Zhang, Dapeng Yu, and Xiaosong Wu.
\newblock {Anomalous Hall effect and magnetic orderings in nanothick
  V$_5$S$_8$}.
\newblock {\em {Phys. Rev. B}}, 96:075402, Aug 2017.

\bibitem{Lee1985}
Patrick~A. Lee and T.~V. Ramakrishnan.
\newblock Disordered electronic systems.
\newblock {\em {Rev. Mod. Phys.}}, 57:287--337, Apr 1985.

\bibitem{Altshuler1980}
B.~L. Altshuler, D.~Khmel'nitzkii, A.~I. Larkin, and P.~A. Lee.
\newblock {Magnetoresistance and Hall effect in a disordered two-dimensional
  electron gas}.
\newblock {\em {Phys. Rev. B}}, 22:5142--5153, Dec 1980.

\bibitem{Kondo1964}
J.~Kondo.
\newblock {Resistance Minimum in Dilute Magnetic Alloys}.
\newblock {\em Progress of Theoretical Physics}, 32(1):37--49, jul 1964.

\bibitem{Ghosh2017}
Tanmoy Ghosh, Takashi Fukuda, Tomoyuki Kakeshita, S.~N. Kaul, and P.~K.
  Mukhopadhyay.
\newblock Concomitant antiferromagnetic transition and disorder-induced weak
  localization in an interacting electron system.
\newblock {\em Physical Review B}, 95(14):140401, apr 2017.

\bibitem{Lu2015}
Hai-Zhou Lu and Shun-Qing Shen.
\newblock {Weak antilocalization and localization in disordered and interacting
  Weyl semimetals}.
\newblock {\em Physical Review B}, 92(3):035203, July 2015.

\bibitem{Lu2014}
Hai-Zhou Lu and Shun-Qing Shen.
\newblock Weak localization and weak anti-localization in topological
  insulators.
\newblock In Henri-Jean Drouhin, Jean-Eric Wegrowe, and Manijeh Razeghi,
  editors, {\em {SPIE} Proceedings}. {SPIE}, aug 2014.

\bibitem{Niu2016}
Wei Niu, Ming Gao, Xuefeng Wang, Fengqi Song, Jun Du, Xinran Wang, Yongbing Xu,
  and Rong Zhang.
\newblock {Evidence of weak localization in quantum interference effects
  observed in epitaxial La$_{0.7}$Sr0.3MnO3 ultrathin films}.
\newblock {\em Scientific Reports}, 6(1):26081, may 2016.

\bibitem{Sett2017}
Shaili Sett, K.~Das, and A.~K. Raychaudhuri.
\newblock Weak localization and the approach to metal{\textendash}insulator
  transition in single crystalline germanium nanowires.
\newblock {\em Journal of Physics: Condensed Matter}, 29(11):115301, feb 2017.

\bibitem{Sasmal2020}
Souvik Sasmal, Rajib Mondal, Ruta Kulkarni, Arumugam Thamizhavel, and Bahadur
  Singh.
\newblock Magnetotransport properties of noncentrosymmetric {CaAgBi} single
  crystal.
\newblock {\em Journal of Physics: Condensed Matter}, 32(33):335701, may 2020.

\bibitem{Laha2021}
Antu Laha, Ratnadwip Singha, Sougata Mardanya, Bahadur Singh, Amit Agarwal,
  Prabhat Mandal, and Z.~Hossain.
\newblock {Topological Hall effect in the antiferromagnetic Dirac semimetal
  {EuAgAs}}.
\newblock {\em Physical Review B}, 103(24):l241112, jun 2021.

\bibitem{Katayama1967}
Yoshifumi Katayama and Shoji Tanaka.
\newblock {Resistance Anomaly and Negative Magnetoresistance in $n$-Type InSb
  at Very Low Temperatures}.
\newblock {\em {Phys. Rev.}}, 153:873--882, Jan 1967.

\bibitem{Hamann1967}
D.~R. Hamann.
\newblock {New Solution for Exchange Scattering in Dilute Alloys}.
\newblock {\em {Phys. Rev.}}, 158:570--580, Jun 1967.

\bibitem{Kondo1999}
S.~Kondo, D.~C. Johnston, and L.~L. Miller.
\newblock Synthesis, characterization, and magnetic susceptibility of the
  heavy-fermion transition-metal oxide {LiV$_2$O$_4$}.
\newblock {\em Physical Review B}, 59:2609--2626, Jan 1999.

\bibitem{Cao2017}
Qiang Cao, Frank~F. Yun, Lina Sang, Feixiang Xiang, Guolei Liu, and Xiaolin
  Wang.
\newblock Defect introduced paramagnetism and weak localization in
  two-dimensional metal {VSe$_2$}.
\newblock {\em Nanotechnology}, 28(47):475703, oct 2017.

\bibitem{Hien-Hoang2021}
Van Hien-Hoang, Nak-Kwan Chung, and Heon-Jung Kim.
\newblock {Electrical transport properties and Kondo effect in
  La$_{1-x}$Pr$_{x}$NiO$_{3-\delta}$ thin films}.
\newblock {\em Scientific Reports}, 11(1):5391, mar 2021.

\bibitem{Yin2020}
Junjie Yin, Changwei Wu, Lisi Li, Jia Yu, Hualei Sun, Bing Shen, Benjamin~A.
  Frandsen, Dao-Xin Yao, and Meng Wang.
\newblock {Large negative magnetoresistance in the antiferromagnetic rare-earth
  dichalcogenide EuTe$_2$}.
\newblock {\em Phys. Rev. Materials}, 4:013405, Jan 2020.

\bibitem{Qi2022}
Bao-Tao Qi, Jun-Jie Guo, Ying qing Miao, Mian zeng Zhong, Bo~Li, Zi~yan Luo,
  Xi~guang Wang, Yao zhuang Nie, Qing lin Xia, and Guang hua Guo.
\newblock {Abnormal Magnetoresistance Transport Properties of van der Waals
  Antiferromagnetic FeNbTe$_2$}.
\newblock {\em Frontiers in Physics}, 10:851838, apr 2022.

\bibitem{Bai2019}
Wei Bai, Zhongqiang Hu, Sheng Wang, Yang Hua, Zhe Sun, Chong Xiao, and Yi~Xie.
\newblock {Intrinsic Negative Magnetoresistance in Van Der Waals {FeNbTe$_2$}
  Single Crystals}.
\newblock {\em Advanced Materials}, 31(18):1900246, mar 2019.

\bibitem{Jovanovifmmodecuteclseci2010}
V.~P. Jovanovi\ifmmode~\acute{c}\else \'{c}\fi{}, L.~Fruchter, Z.~Z. Li, and
  H.~Raffy.
\newblock Anisotropy of the in-plane angular magnetoresistance of
  electron-doped {Sr$_{1-x}$LA$_x$CuO$_2$} thin films.
\newblock {\em {Phys. Rev. B}}, 81:134520, Apr 2010.

\bibitem{Singha2018}
Ratnadwip Singha, Arnab Pariari, Gaurav~Kumar Gupta, Tanmoy Das, and Prabhat
  Mandal.
\newblock {Probing the Fermi surface and magnetotransport properties of
  MoAs$_{2}$}.
\newblock {\em {Phys. Rev. B}}, 97:155120, Apr 2018.

\bibitem{Wang2021a}
Junjie Wang, Jun Deng, Xiaowei Liang, Guoying Gao, Tianping Ying, Shangjie
  Tian, Hechang Lei, Yanpeng Song, Xu~Chen, Jian-gang Guo, and Xiaolong Chen.
\newblock {Spin-flip-driven giant magnetotransport in A-type antiferromagnet
  {NaCrTe$_2$}}.
\newblock {\em Phys. Rev. Materials}, 5:L091401, Sep 2021.

\bibitem{McGuire1975}
T.~McGuire and R.~Potter.
\newblock Anisotropic magnetoresistance in ferromagnetic 3d alloys.
\newblock {\em {IEEE} Transactions on Magnetics}, 11(4):1018--1038, jul 1975.

\end{thebibliography}





% \end{thebibliography}

\end{document}